\documentclass{ws-ijmpa}

\begin{document}

\markboth{A. Rybarska, W. Sch\"afer, A. Szczurek}
{Exclusive $J/\psi$ photoproduction in $pp$ and $p \bar p$ collisions} 

\catchline{}{}{}{}{}
%
%%%%%%%%%%%%%%%%%%%%%%%%%%%%%%%%%%%%%%%%%%%%%%%%%%%%%%%%%%%%%%%%%%%%

\title{EXCLUSIVE $J/\psi$ PHOTOPRODUCTION IN $pp$ AND $p\bar p$ COLLISIONS
}

\author{ANNA RYBARSKA$^\star$, WOLFGANG SCH\"AFER$^{\star,}$\footnote{E-mail:
Wolfgang.Schafer@ifj.edu.pl}, ANTONI SZCZUREK$^{\star,\%}$
}

\address{$^\star$ Institute for Nuclear Physics, PAN, 
PL-31-342 Krak\'ow, Poland\\
$^\%$ University of Rzesz\'ow, PL-35-959 Rzesz\'ow, Poland} 

\maketitle

%\begin{history}
%\received{Day Month Year}
%\revised{Day Month Year}
%\end{history}

\begin{abstract}
We report on a detailed investigation of exclusive $J/\psi$
production in proton--proton and proton--antiproton collisions.
Predictions for Tevatron and LHC energies are included.

\end{abstract}

\ccode{PACS numbers: 13.87.Ce, 13.60.Le, 13.85.Lg}

\section{Introduction}	
Diffractive production of vector mesons has been a major subject
at HERA, and the large body of data obtained serves to 
elucidate the physics of the QCD--Pomeron and/or the small--$x$ 
gluon density in the proton \cite{INS06}. 
For a review at this workshop, see
\cite{Marage}. While data--taking at HERA has come to an end, hadron
colliders, such as Tevatron and LHC provide opportunities to study
photoproduction processes at still higher energies than those 
covered by HERA.
The exciting experimental possibilities at Tevatron
at LHC have been presented at this workshop
in the plenary talk \cite{Pinfold}. 
Besides the photoproduction mechanism of exclusive $J/\psi$--production in 
hadronic collisions, there could also
exist a purely hadronic mechanism, where the $J/\psi$ production
proceeds through the Pomeron--Odderon fusion. 
From this point of view a detailed study of the photoproduction
mechanism, if possible including the effects of absorptive
corrections is much desired. 
The results presented here are based on ref. \cite{SS07} and ongoing
work with A. Rybarska. For an analogous study of $\Upsilon$ production,
see \cite{RSS}.

\section{Results and Conclusions}

\begin{figure}[pb]
\centerline{\psfig{file=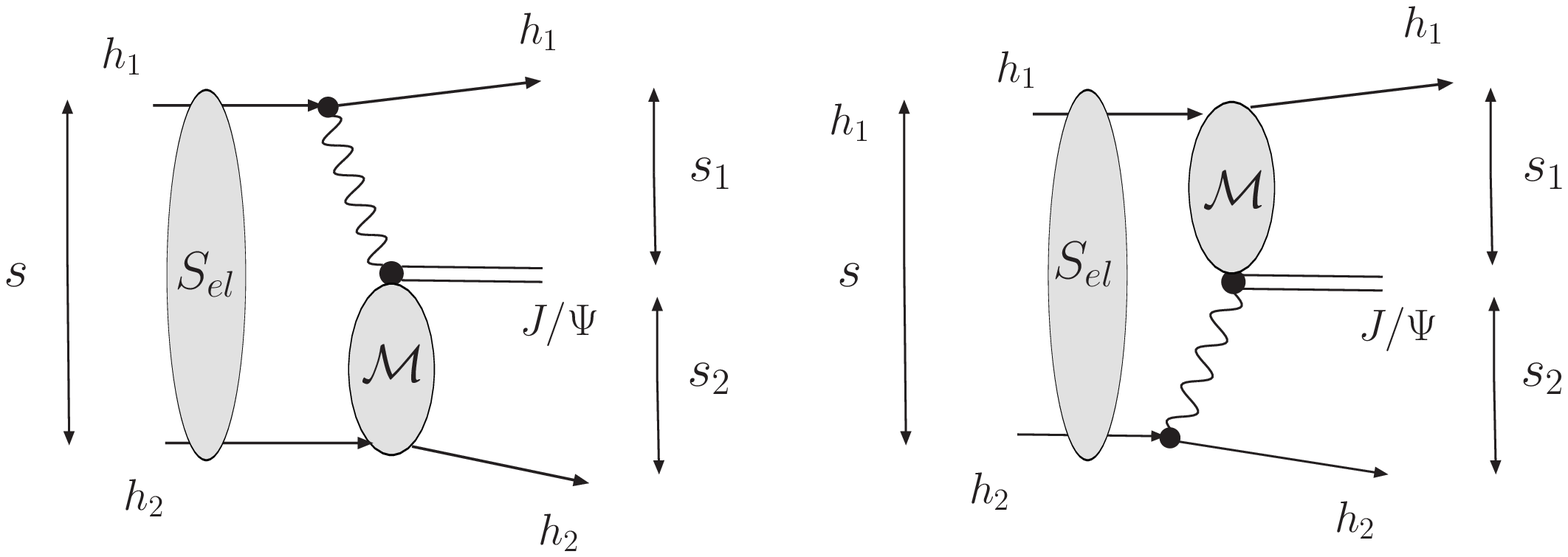,width=12cm}}
\vspace*{8pt}
\caption{A diagrammatic representation of the exclusive $J/\psi$--production mechanism 
in hadron--hadron scattering. Here the blob labeled ${\cal{M}}$ corresponds to the
photoproduction amplitude of $J/\psi$ on hadron $h_{1,2}$. The blob $S_{el}$ stands
for the absorptive correction. Some kinematic variables are indicated. \label{f1}}
\end{figure}

\begin{figure}[pb]
\centerline{\psfig{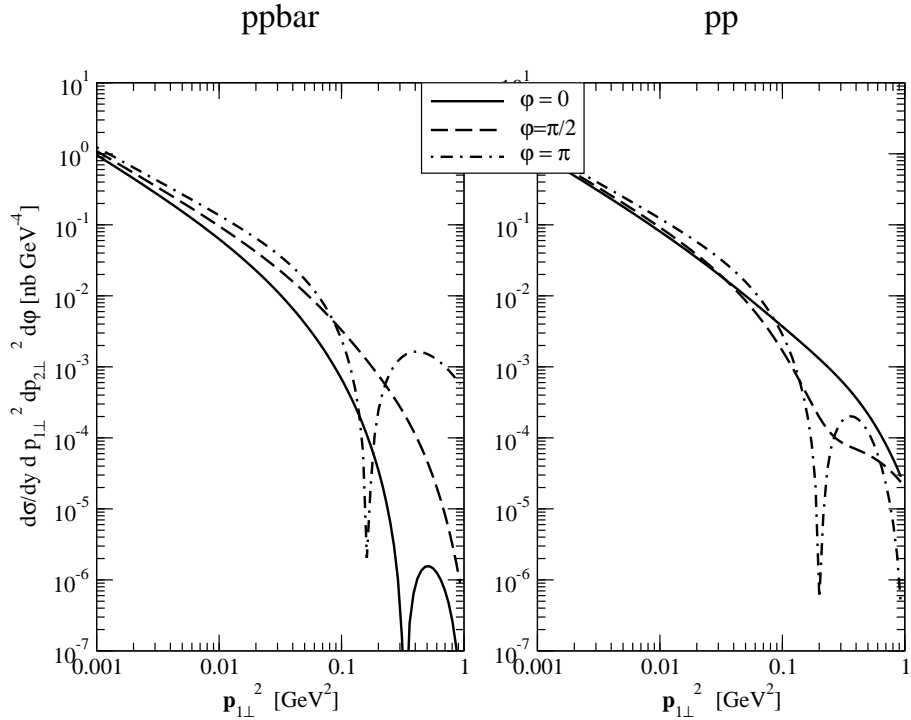}}
\vspace*{8pt}
\caption{
{\bf{Left}}: Fully differential cross section as a function of ${\bf{p}}^2_1$ at $y = 0$
and ${\bf{p}}_2^2 = 1$ GeV$^2$ for $p \bar p$ collisions at $W = 1960$ GeV.
{\bf{Right}}: the same for $pp$ collisions.
\label{f2}}
\end{figure}

\begin{figure}[t]
\begin{center}
\begin{minipage}[t]{0.48\textwidth}
\centerline{\epsfysize 6.4 cm
\epsfbox{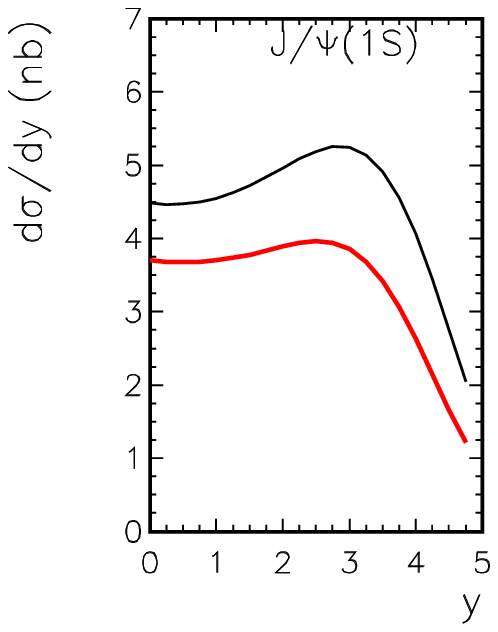}}
\end{minipage} \hfill
\begin{minipage}[t]{0.48\textwidth}
\centerline{\epsfysize 6.4 cm
\epsfbox{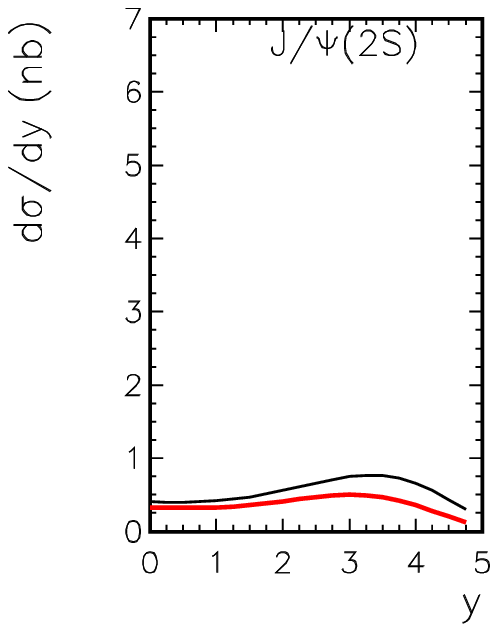}}
\end{minipage}
\vspace{-0.5cm}
\caption{\label{f3} 
{\bf{Left}}: $d\sigma/dy$ at $W=1960$ GeV for $J/\psi$--production.
The upper curve is without absorptive corrections, while the lower curve 
includes them.
{\bf{Right}}: the same for $\psi' (2S)$.
}
\end{center}
\end{figure}

\begin{figure}[t]
\begin{center}
\begin{minipage}[t]{0.48\textwidth}
\centerline{\epsfysize 6.4 cm
\epsfbox{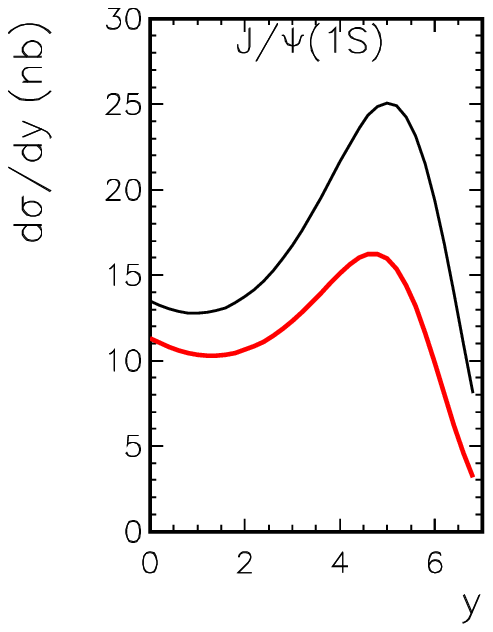}}
\end{minipage} \hfill
\begin{minipage}[t]{0.48\textwidth}
\centerline{\epsfysize 6.4 cm
\epsfbox{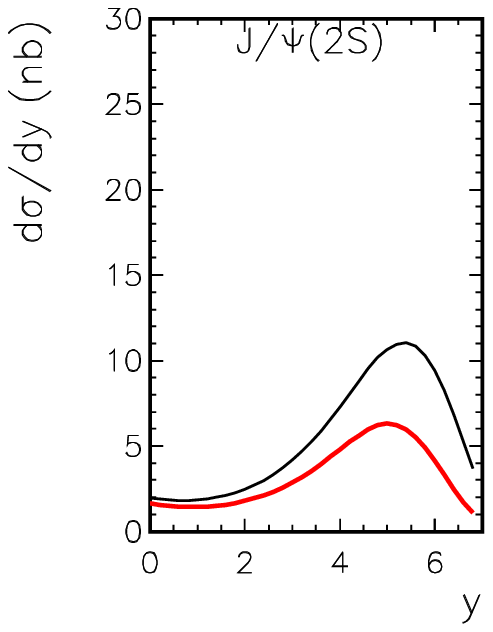}}
\end{minipage}
\vspace{-0.5cm}
\caption{\label{f4} 
{\bf{Left}}: $d\sigma/dy$ at $W=14$ TeV for $J/\psi$--production.
The upper curve is without absorptive corrections, while the lower curve 
includes them.
{\bf{Right}}: the same for $\psi' (2S)$.
}
\end{center}
\end{figure}

The mechanism for exclusive $J/\psi$ production is depicted in fig.(\ref{f1}). 
The exchanged photons have small virtualities, certainly $Q^2 \ll m^2_{J/\psi}$,
and what enters the calculation is the photoproduction amplitude $\gamma p \to J/\psi
p$. Under mild assumptions, this subprocess amplitude can be directly
obtained from the HERA data on $t$--distributions of $J/\psi$'s,
\begin{equation}
{\cal{M}}(\gamma p \to J/\psi p;s,t) \propto (i + \rho) \sqrt{d\sigma(s)/dt|_{t=0}} \exp(B t/2) \, .
\end{equation}
In particular, we assumed that $s$--channel helicity conserving pieces dominate, which
is indeed borne out by experiment, see the detailed review \cite{INS06} 
for references. The second ingredient is the amplitude for elastic scattering of
the incoming (anti--)protons, which enters in the evaluation of 
absorptive/rescattering corrections. We also essentially take it from data. 
Our resulting predictions are then only mildly model--dependent.
Some reservations exist regarding the possible presence of inelastic 
intermediate states in the rescattering diagram, which can however be estimated
by simple enhancement factors.
From the large amount of detailed plots shown 
in our presentation we select a few representative ones:
In fig(\ref{f2}), we show a scan of the fully differential cross section
$d\sigma / dy d{\bf{p}}_1^2 d{\bf{p}}_2^2 d \phi$, at Tevatron energies. 
Here $y$ is the rapidity of the $J/\psi$, 
${\bf{p}}_{1,2}$ are the transverse momenta of the outgoing (anti--)protons, 
and $\phi$ is the azimuthal angle between them. 
The dependence on $\phi$ is caused by the interference of
the two diagrams of fig(\ref{f1}), which enters with a different sign in $pp$ 
and $p \bar p$ collisions. The steep peak at small ${\bf{p}}_1^2$ is naturally due
to the photon pole. The rich dip--bump structure is fairly sensitive to the
precise treatment of absorptive corrections, for more details, see \cite{SS07}.
The plots of figs(\ref{f3}, \ref{f4}) are results not contained in \cite{SS07}.
Here, for the $\gamma p \to J/\psi p$ subprocess we did not use a purely
phenomenological parametrisation, but used a pQCD--based $k_\perp$--factorisation
approach \cite{INS06} to calculate the relevant amplitude in terms of
an unintegrated gluon density of the proton and the light--cone wave function 
of the final state vector meson. This machinery allows us to calculate also 
production of excited vector mesons, for which HERA data are rather sparse 
and insufficient for predictions directly from data. 
In fig(\ref{f3}) we show the rapidity spectra $d\sigma/dy$ at Tevatron energies.
In the left panel we show results for the $J/\psi(1S)$, in the right panel we 
display the cross section for production of the radial excitation $\psi'(2S)$.
Here, the light cone wave functions of $J/\psi, \psi'(2S)$ which we used 
are purely phenomenological
ans\"atze, constrained by the decay width to leptons, as well as the 
orthogonality condition \cite{INS06}. We checked, that our form of the 
wavefunction leads to a good description of the $\psi'/J/\psi$--ratio
measured in photoproduction at HERA. The size of the $\psi'$ production
cross section roughly corresponds to the $\psi'/ J/\psi$--ratio in
photoproduction. In fig(\ref{f4}) we show predictions for LHC--energies.
An accurate treatment of absorption
at LHC energies must go beyond the elastic rescattering included so far,
still our estimates of fig(\ref{f4}) could be of some use. 
The rise of the $y$--distribution from central rapidities to
the peak reflects the Pomeron intercept in the $\gamma p \to J/\psi p$ production
amplitude, and could be further affected by absorptive corrections 
to that subprocess. Already the included absorptive corrections
affect the shape of the rapidity distribution, because
larger $y$ also involves larger photon virtualities, and hence 
shorter distances, where absorption is stronger.
While analysis is in progress at the Tevatron \cite{Pinfold}, measurements of exclusive 
$J/\psi$ at LHC will be very challenging,
prospects are better for $\Upsilon$'s discussed in \cite{RSS}.

\section*{Acknowledgments}

It is a pleasure to thank Jim Pinfold for discussions during MESON2008.
This work was partially supported by the 
Polish Ministry of Science and Higher Education (MNiSW) under contract 1916/B/H03/2008/34.

%\begin{thebibliography}{000} %for 3 digits
%\begin{thebibliography}{00}  %for 2 digits

\end{document}